\documentclass[preprint,12pt]{elsarticle}

\usepackage{amssymb}

\usepackage{amsthm}
\usepackage{amsmath}
\usepackage{epstopdf}
\usepackage{siunitx}
\usepackage{graphicx}
\usepackage{multirow}
\usepackage{multicol}
\usepackage[normalem]{ulem}
\usepackage{lipsum}
\usepackage{subcaption}
\usepackage{xcolor}
\usepackage{psfig}
\usepackage{epsf}

\begin{document}

\begin{frontmatter}

\title{Henry's constant of helium in liquid lead-lithium alloys}

\author{E. Álvarez-Galera$^{a,b}$}
\affiliation{organization={Department of Physics, Polytechnic University of 
            Catalonia-Barcelona Tech},
            addressline={B4-B5 Northern Campus UPC}, 
            city={Barcelona},
            postcode={08034}, 
            state={Catalonia},
            country={Spain}}
            \affiliation{organization={E.T.S.E.I.B, Polytechnic University of Catalonia-Barcelona Tech},
            addressline={Diagonal 647}, 
            city={Barcelona},
            postcode={08028}, 
            state={Catalonia},
            country={Spain}}

\author{D. Laria$^{c}$}
\affiliation{organization={Institute of Chemistry, Eötvös Loránd University},
            addressline={P.O. Box 32}, 
            city={Budapest},
            postcode={112 H-1518}, 
            country={Hungary}}
\author{F.Mazzanti$^a$}
\author{L. Batet$^b$}
\author{J. Martí$^{a,d}$}
\affiliation{Corresponding author, email: jordi.marti@upc.edu}

\begin{abstract}
    The solubility of helium in liquid metals is a knowledge of fundamental importance in the design of the future nuclear fusion reactors, since the formation of helium bubbles inside the breeding blankets of the reactors can be a threat to the durability of the devices and,  more importantly, to the efficiency of tritium's recovery.  In the present work we report a detailed set of calculations of the solubility of helium in lead and lead-lithium alloys.  A series of molecular dynamics simulations have been combined with a classical perturbative procedure able to compute the free energy of insertion of a helium atom inside a liquid metal bath, directly related to the solubility of helium. As the most important case, the concentration of the eutectic solution has been explored in full detail. We have found that 
solubility of helium in pure lithium is lower than in pure lead,  predicting a value at the eutectic state (16\%Li-84\%Pb at 508~K) of 
about $5\times 10^{-16}$~Pa$^{-1}$.  The observed trend indicates that solubilties rise with increasing temperatures.
\end{abstract}

%%Graphical abstract
\begin{graphicalabstract}
{\color{black}
\includegraphics[width=0.5\textwidth]{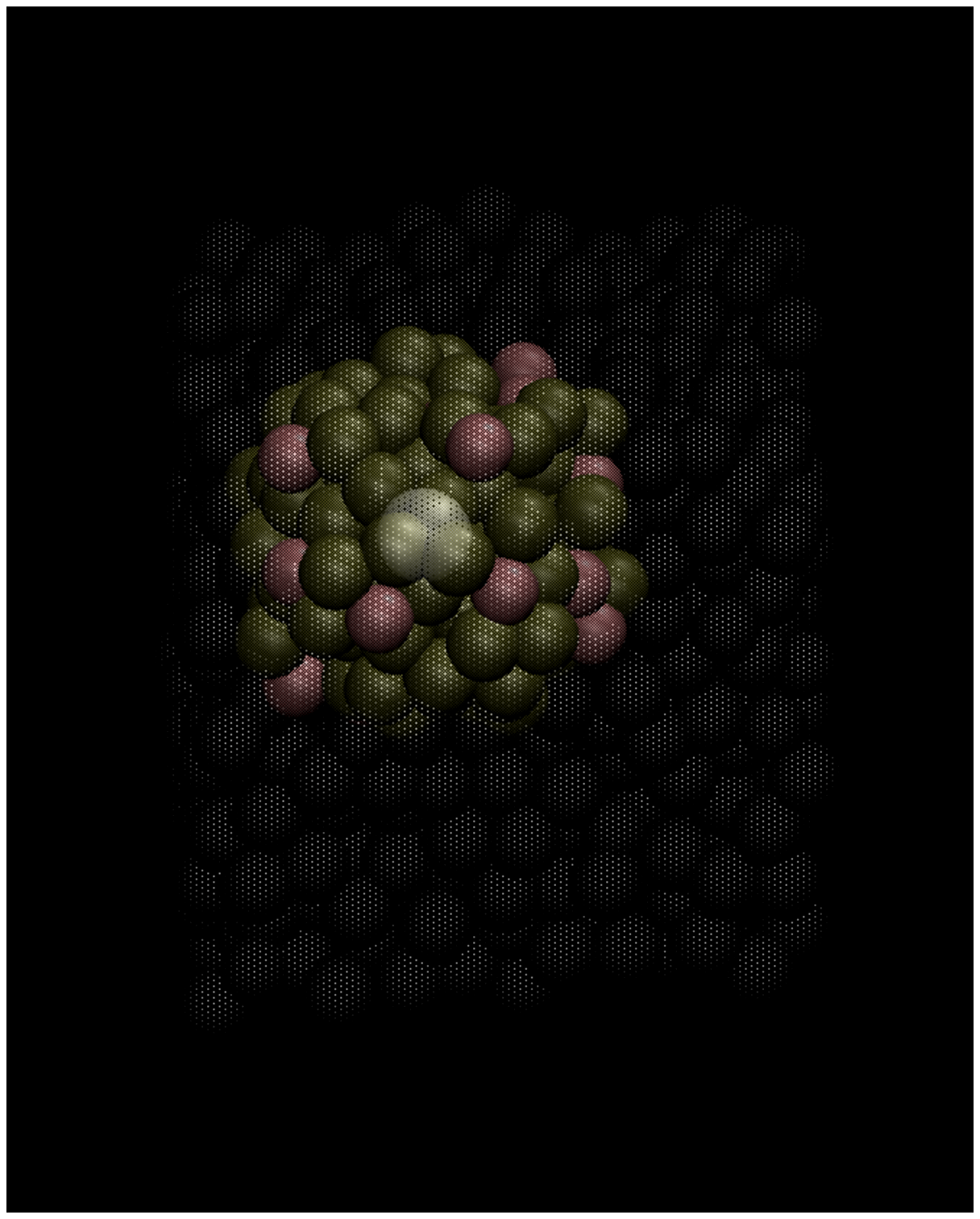} \\
Pb atoms: green;
Li atoms: red;
He atom: white
\\
Only those atoms within a cutoff of 10~\AA \  have been colored.}
\end{graphicalabstract}

%%Research highlights
\begin{highlights}
\color{black}
\item Highlight 1.  The formation of helium bubbles can be an important issue in the design of future nuclear fusion reactors.
\item Highlight 2.  The predicted solubility of helium at the eutectic point (16\%Li-84\%Pb at 508~K) is of $5\times 10^{-16}$~Pa$^{-1}$.
\item Highlight 3.  The solubility of helium rises with increasing temperatures.
\end{highlights}


\end{frontmatter}

%% \linenumbers

\definecolor{green4}{RGB}{0,128,0}
\definecolor{orange}{RGB}{255,165,0}
\definecolor{violet}{RGB}{116,1,113}

%% main text
\section{Introduction}
\label{intro}
    
The determination of the solubility of helium in liquid metals (LM) is relevant from the physical and technological points of view, mainly because helium plays a relevant role in some of the fundamental processes that take place in nuclear reactors.\cite{reed1970solubility,borgstedt2001iupac} Liquid sodium, potassium, and sodium-potassium mixtures have proven to be adequate materials to act both as heat transfer media and as coolants in the pioneering fast-breeder fission nuclear reactors. 
Alkali metals, though, react chemically with oxygen or carbon dioxide in air, so it is important to provide a protective blanket which normally consists of none-reactive gases such as helium or argon, for instance.  However, the use of these combinations presents some drawbacks because accumulations of inert gases may lead to interference of heat transfer or to changes of the reactivity in the core of the reactors.  Consequently,  properties such as the solubility and temperature dependence or the diffusivity of noble gases under gradient concentrations are key ingredients that should be considered in appropriate and efficient designs.  More than fifty years ago, a pioneering work by Epstein\cite{epstein1951solubility} estimated the solubility of helium in sodium to be of the order of $10^{-10}$~cm$^3$ of helium per cm$^3$ of sodium at temperatures close to  200~$^{\circ}$C.  Given its low solubility, the potential nucleation of helium in LM may be a relevant engineering problem: the main breeding reaction in a fusion reactor is n+$^6$Li$\rightarrow$ T+He+ 4.8 MeV, with a He production as large (mol to mol) as the production of tritium.

The solubility of monoatomic gases in a liquid is normally determined by Henry's law: $x = \frac{P}{K_{\rm H}}$, where $x$ is
the atomic fraction of saturated gas, $P$ the partial pressure of the gas over the liquid, and $K_{\rm H}$ the volatility constant of Henry's law (hereafter Henry's constant), which is inverse to the constant of solubility.  Hence, the estimation of $K_{\rm H}$ provides us a threshold condition for helium nucleation.  The experimental measurements of $K_{\rm H}$ are technically complex, the most relevant measures being those by Slotnick et al.\cite{slotnick1965solubility} who measured the solubility of helium in lithium and potassium, whereas Thormeier\cite{thormeier1970solubility} measured the solubility of argon and helium in liquid sodium, 
and Veleckis et al.\cite{veleckis1971solubility} reported the solubility of helium and argon in liquid sodium.  To the best of our knowledge, 
no experimental references on the solubility of helium in the lead-lithium eutectic (LLE) have been reported yet.  The LLE (with $15.7-17\%$ of lithium,\cite{malang1995comparison,sedano2007helium,de2008lead,kordavc2017helium,khairulin2017volumetric}) is of 
great practical importance, since most of the breeding blanket designs in fusion reactors use it,  and where the main 
role of lead is to reduce the extreme reactivity of pure lithium and to act as a neutron multiplier.  Experimental measurements by Tazhibayeva\cite{tazhibayeva2014interaction} and Kulsartov et al.\cite{kulsartov2019study} described the process of generation 
and release of tritium and helium from the LLE, thus indicating that the flow of helium from the eutectic’s surface is linearly dependent 
on its bulk concentration.  Predictions of the solubility of helium in the LLE indicate values between 1.14$\times$10$^{-17}$ and 
1.35$\times$10$^{-15}$ Pa$^{-1}$,\cite{sedano2022solubility} whereas a new experimental setup for its direct measurement 
has been recently proposed.\cite{kekrt2023concept}
    
In a previous work\cite{alvarez2023nucleation} we established a set of force fields capable of reproducing, by molecular dynamics (MD), many relevant experimental properties of helium in liquid lithium, with the aim of analyzing helium nucleation at the atomic level.  
Later on,\cite{alvarez2024henry} we considered such family of models to estimate the solubility of helium in lithium, sodium, potassium, rubidium and cesium, and found an excellent agreement with the solubility data reported by Slotnick,\cite{slotnick1965solubility} Thormeier\cite{thormeier1970solubility} and Veleckis\cite{veleckis1971solubility} referred above, together with a fair validation of potential models through the direct comparison to experimental structure factors in all cases.  We used the so-called cavity method to compute Henry's constants, tested successfully for naphthalene, phenanthrene, calcite, aragonite and cholesterol in water.\cite{li2017computational,li2018computational,wand2018addressing,espinosa2018calculation}. The cavity-creation scheme allowed us to compute $K_{\rm H}$ from the change of free energy when one helium atom is inserted inside a bath of LM atoms.  
In the present work, and in order to perform a systematic study,  we report the solubility of helium in the lead-lithium mixture within a wide range of concentrations, from those at low lead content towards the LLE around 16$\%$ of lithium and on.  
  
The work is organized as follows: in Section~\ref{methods} we describe the potential models, provide technical details of the simulations and sketch the gross features of the cavity method.  In Section~\ref{theory} we describe the relationship between the cavity method and the excess chemical potential.  In Section~\ref{results} we report and analyze Henry's constant for helium in both pure lead and pure lithium, and in lead-lithium, paying special attention to the case of LLE.  Finally, in Section~\ref{conclusions} we give some concluding remarks and summarize the work.

\newpage

\section{Methods and Theory}
\label{methods}
\subsection{Model Hamiltonian}
\label{model}
 
We model Li and Pb atoms using a massive point-like description of those, which interact via embedded-atom-model (EAM) interactions

 \begin{eqnarray}
          V_0 (\left\{{\bf r}^{\alpha} \right\})  &=& \sum_\alpha \sum_{i}^{N_\alpha} \left[\Phi_{\rm \alpha} (\psi_i^\alpha) +\frac{1}{2} \sum_{\gamma} \sum_j^{N_\gamma}  \phi_{\alpha \gamma}(r_{ij}^{\alpha \gamma})\right]
        \ \ . 
    \end{eqnarray}
In the last expression, 
$\alpha$ and $\gamma$ denote the species (either Li or Pb) of the $i$-th and $j$-th atoms, respectively, while 
$N_\alpha$ denotes the number of atoms of species $\alpha$, ${\bf r}_i^\alpha$ denotes   
the position of the $i$th atom of species $\alpha$
and $r_{ij}^{\alpha\gamma} = | {\bf r}_i^\alpha - {\bf r}_j^\gamma|$ is the distance between two given atoms $i,j$ of species $\alpha,\gamma$, respectively.  The collective variables $\psi_i^{\alpha}$ (effective electronic density at site $i$) are defined as
    \begin{eqnarray}
        \psi_i^{\alpha} =
        \sum_{\gamma}  p_1^\gamma 
        \sum_{j=1}^{N_\gamma} \exp{\left( -p_2^\gamma \ r_{ij}^{\alpha\gamma} \right)}  (1-    \delta_{\alpha\gamma}\delta_{ij}),
    \end{eqnarray}
    
where $p_k^\gamma$ ($k=1,2$) are 2 constant parameters that depend on the type of neighboring atom $j$ and 
$\delta_{mn}$ are Kronecker's deltas (full details can be found in the original references by Belaschenko\cite{belashchenko2012embedded,belashchenko2012electron}.  Our parametrization of the latter EAM potential is 
similar to the one proposed by Al-Awad et al.,\cite{al2023parametrization} which proved to be adequate to describe density 
fluctuations in different LiPb mixtures. 

On the other hand, our parametrization of He-solvent interactions relies on a pair-decomposable model. 
As such, the expression of the solute-solvent contribution to the potential energy reads:
\begin{equation}
    V_{\rm He -S} ({\bf r}^{\rm He}, \{{\bf r}^{\alpha}\}) =
     \sum_\alpha \sum_{i}^{N_{\alpha}} \phi_{{\rm He}-\alpha} 
     (| {\bf r}_i^{\alpha} - {\bf r}_{\rm He} |) \; .
\end{equation}
    
For $\phi_{\rm He-Li} (r)$, we use the  Toennies-Tang-Sheng (TTS) parametrization,\cite{sheng2021development} 
whereas the $\phi_{\rm He-Pb} (r)$ interactions are modeled using the {\it ab initio} expression proposed by Sladek et al. in Ref.~\cite{sladek2014ab}.
Remarkably, the main characteristics of the latter potentials -- which play quite a relevant role in capturing the main characteristics of the Henry's constant -- can be cast in terms of a steep repulsive branch, crossing zero at about $r=5.35$~\AA \ and $r=4.15$~\AA \ for Li and Pb, respectively, followed by a milder attractive tail at long distances, with typical energy scales much smaller than the magnitudes of thermal energy that we will investigate in this study. 
      
\subsection{Technical details}

\begin{figure}
    \centering
    \includegraphics[width=1\linewidth]{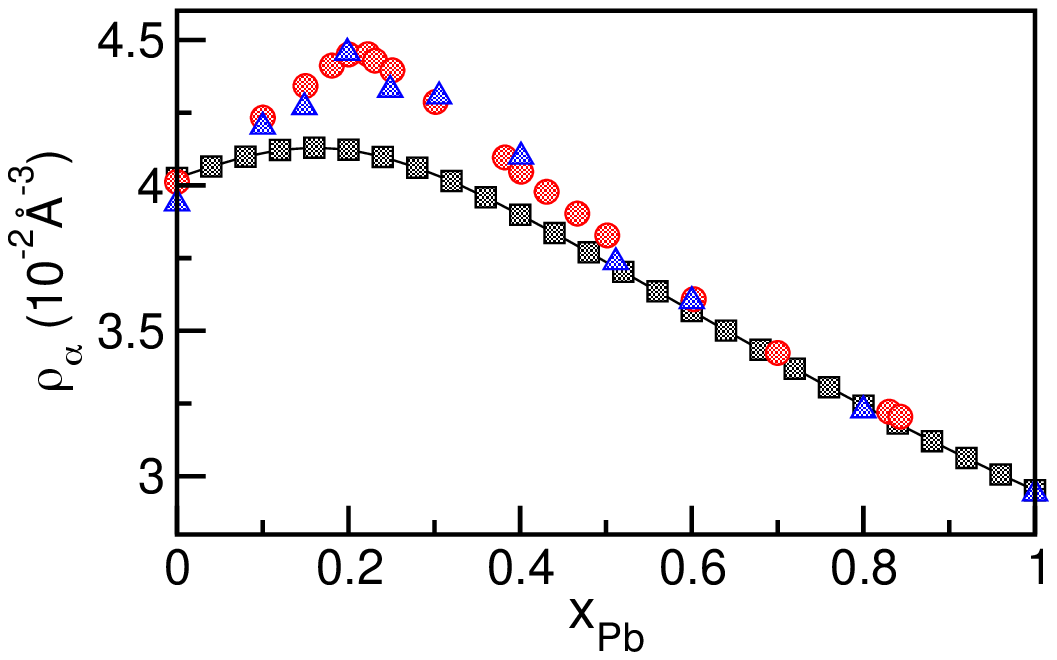}
    \caption{Density dependence on the Pb atomic fraction of different melts (grey squares).
    Also shown are experimental values reported in Ref.~\cite{khairulin2017volumetric} (red circles) and 
    Ref.~\cite{saar1987calculation}
    (blue triangles).}
    \label{fig:density_vs_x}
\end{figure}

We collect statistics from molecular dynamic trajectories evaluated in the isothermal-isobaric (NPT) ensemble, using the {\it LAMMPS} package.\cite{LAMMPS} Appropriate temperature and pressure controls are obtained by coupling the systems to Nos\'e-Hoover thermostats and barostats.  In all cases, the pressure of the barostat is fixed at 1 $\times 10^{5}$ Pa;
resulting in simulated average densities within $90\%$ of the experimental values,\cite{khairulin2017volumetric, saar1987calculation} as shown in Fig.~\ref{fig:density_vs_x}.

All systems comprised a total of $N_{\rm Li}+N_{\rm Pb}= 1024 $, allowing the computation of density fluctuations up to typical distances of the order of $\sim 15$~\AA. 

\begin{figure}
    \centering
   \includegraphics[width=1\linewidth]{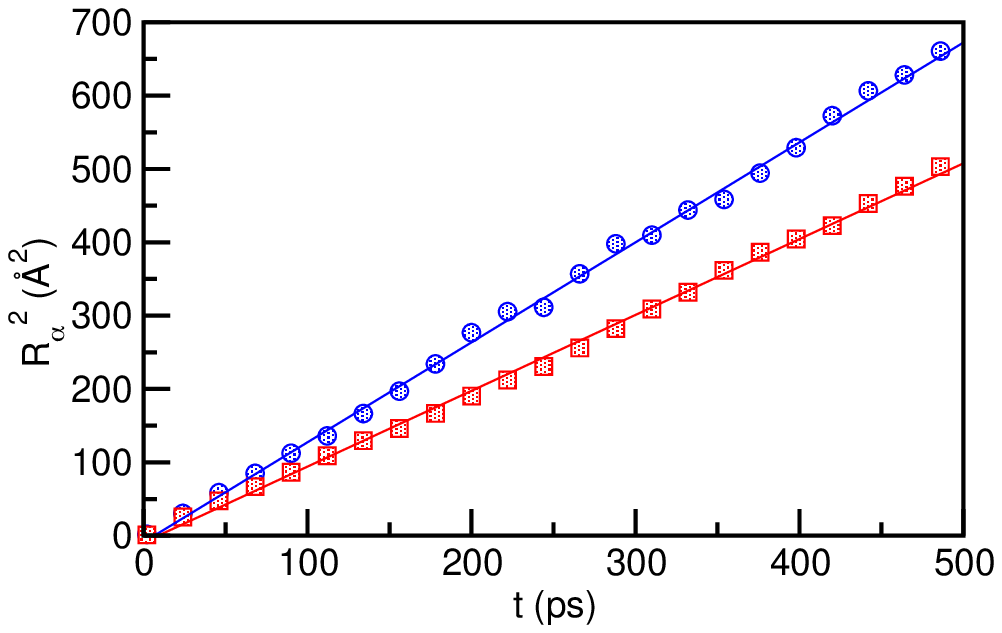}
    \caption{Mean square displacements of Li (blue circles) and Pb (red squares) corresponding to the eutectic mixture at 508 K. Lines are an aid to the eye.}
    \label{fig:MSDeut}
\end{figure}

We have verified that the mixtures presented liquid-like dynamical characteristics, at all the thermodynamic conditions considered, as reflected by the long-time linear behavior of the corresponding mean square displacements,\cite{hansen2013theory}

\begin{equation} \label{msd}
{\cal R}^2_\alpha(t) = \frac{1}{N_\alpha}\sum_{i=1}^{N\alpha}\langle 
|{\bf r}_\alpha^i(0) -{\bf r}^i_\alpha(t)|^2\rangle .\ \ 
\end{equation}
as shown in Fig.~\ref{fig:MSDeut} for a representative example.

\begin{figure}
    \centering
    \includegraphics[width=1\linewidth]{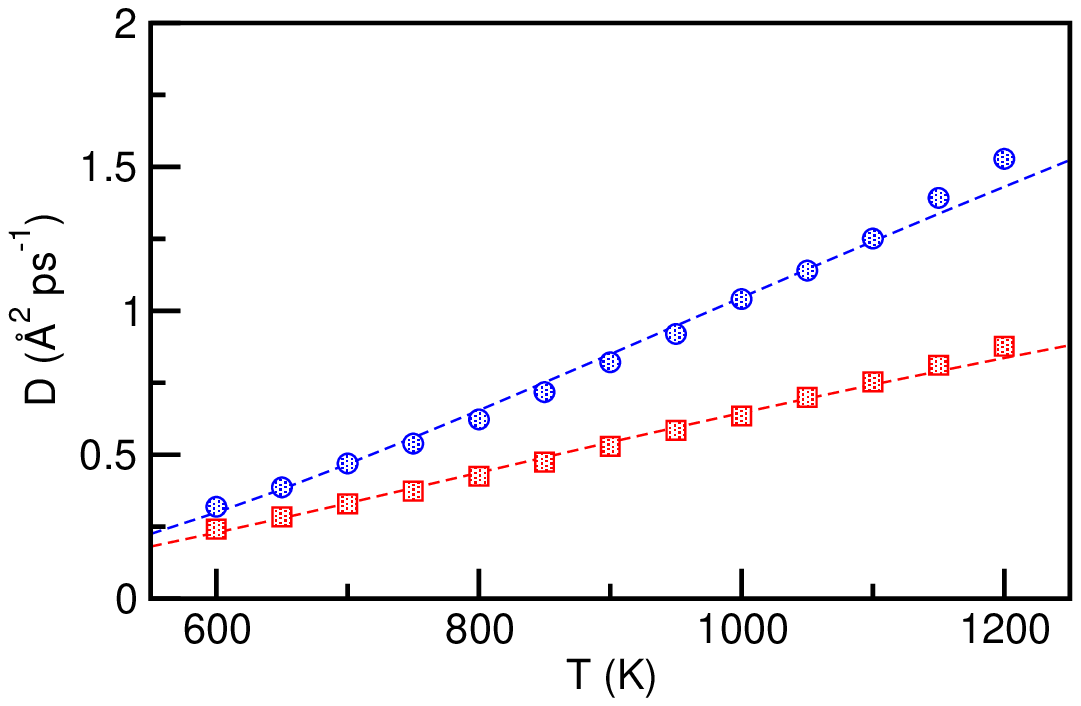}
    \caption{Temperature dependence of the diffusion coefficients of Li (blue circles) and Pb (red squares) for the mixtures at the eutectic concentration.  The lines are an aid to the eye.}
       \label{fig:dif_vs_T}
\end{figure}

For the eutectic particular case, our simulations predict diffusion coefficients of $D_{\rm Li}=0.2271 \pm 0.0006$~\AA$^2$ps$^{-1}$ and $D_{\rm Pb}=0.1723\pm0.0005~$\AA$^2$ps$^{-1}$. To the best of our knowledge,  in the literature there is no reported experimental information of these magnitudes to gauge the accuracy of our results.  Then,  as additional information,  in Fig.~\ref{fig:dif_vs_T} we show the temperature dependence of the Li and Pb diffusion coefficients for the eutectic mixture, 
during the first series, calculated from the slope of Eq.~\ref{msd}.

\begin{figure}
    \centering
    \includegraphics[width=1\linewidth]{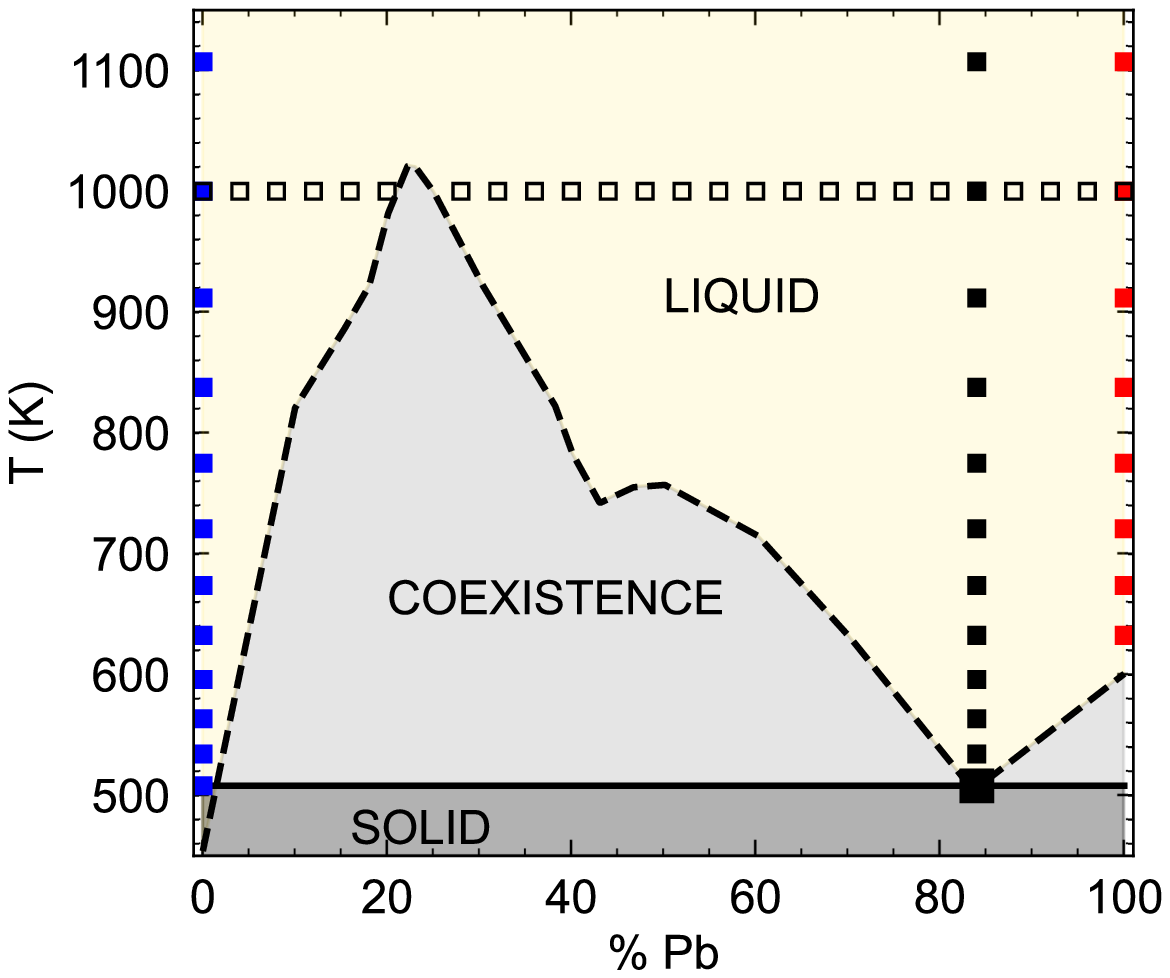}
    \caption{Lead-lithium phase diagram.
        The black, blue and red squares correspond to states for the eutectic mixture, Li and Pb (respectively) investigated at different temperatures.  The empty squares represents the 1000-K isotherm.
    The yellow area contains all the states at which the solvent must be molten, while grey regions have not been investigated. Light gray regions correspond to (experimental) liquid-solid phase coexistence areas, while the dark gray region corresponds to the solid phase.
     The light gray regions correspond to liquid-solid phase coexistence areas.
     }
    \label{fig:DF2}
\end{figure}

We have covered a wide range of solvent compositions and temperatures. We started by examining 16$\%$Li-84$\%$Pb eutectic mixture at $T=508$ K. To examine the temperature and concentration dependences of the Henry's constant, two additional series of simulation runs were performed.  In the first series, we reached a temperature of $T=1200$ K, keeping the eutectic concentration fixed; in the second one, we explored the dependence of the solubility concentration along the 1000~K-isotherm corresponding to the minimum temperature 
at which any Pb-Li mixture is at the liquid phase.  In Fig.~\ref{fig:DF2} we present a scheme of the phase diagram of the mixtures along with the set of thermodynamic states explored in the two series of runs previously described. 

\subsection{Henry's constant} \label{theory}
From a microscopic perspective, the Henry's constant of He, $K_H$, is intimately related to the value of the excess chemical potential 
of the solute at infinite dilution in the different solvents, $\mu_{ex}$, i.e.\cite{murad2000simple}
\begin{equation}
    \label{kh}
    K_{H}=\beta^{-1}\rho_S \ e^{\beta\mu_{ex}} \; ,
\end{equation}
where $\beta$ represents the inverse temperature of the system and $\rho_S$ corresponds  to the global density of an $\alpha$ tagged solvent.  The excess chemical potential represents the reversible work involved in the insertion of a single solute {\color{black}} atom within the solvent phase.  Such reversible work can be computed using one-step (Widom) methods, either Monte Carlo or MD, from the following average value:\cite{jcp2808-63}
\begin{equation}
\label{mu}
\mu_{ex}= -\beta^{-1}\ln \langle e^{-\beta V_{\rm He-\alpha}}\rangle_0^{\rm NVT} \ \ ;
\end{equation}
where $\langle \dots\rangle_0$ stands for a statistical average computed over all solvent configurations,
computed from canonical (NVT) trajectories.  The subscript 0 indicates that averages are calculated respect to the reference 
system.  For NPT simulations the correct expression reads:
\begin{equation}
    \label{npt1}
    \mu_{ex}=-\beta^{-1}\ln \left
    [
    \frac{\langle Ve^{-\beta V_{\rm He-\alpha}}\rangle_0^{\rm NPT}}{\langle V\rangle_0^{\rm NPT}}\right] \ \ .  
\end{equation}

From an operational perspective, standard perturbation theory (see Ref.\cite{chandler1987introduction}) provides an appropriate alternative to compute $\mu_{ex}$ via the following expression:
\begin{equation}
    \label{eq1}
    \mu_{ex}=\int_0^\lambda {\rm d}\lambda \ 
    \Biggl \langle
    \frac{\partial H(\lambda)}{\partial \lambda}
    \Biggr \rangle
    _{H(\lambda)} = \int_0^\lambda {\rm d}\lambda \ 
     \langle V_{\rm He-\alpha}\rangle
    _{H(\lambda)} \ \ . {\color{black},}
\end{equation}
often referred as {\it coupling parameter}\cite{chandler1987introduction} or {\it thermodynamic integration}\cite{frenkel2001understanding, anwar2005robust, carruthers2023prediction} approach.
In the last equation, $\langle \cdots\rangle_{H(\lambda)}$ denotes a statistical average controlled by a scaled Hamiltonian with potential energy $V(\lambda)= V_0+\lambda V_{\rm He-\alpha} $. 
The first equality is a general result, whereas the second one is a consequence of the linear $\lambda$-coupling in the solute-solvent potential. 

The characteristics of $V_{\rm He-\alpha}$\cite{sheng2021development, sladek2014ab} prevents a direct implementation of Eq.~(\ref{eq1}) since it presents a well documented repulsive singularity at short distances.\cite{alvarez2024henry}
A valid alternative route was proposed by Li et al.\cite{li2017computational} in which the insertion of the guest solute consisted of three stages.  During the first one, hereafter referred to as the {\it growth} stage,  the chemical potential 
of the solute is computed using a solute-solvent interactions which is gradually 
turned on via a non divergent, repulsive potential of the form:
\begin{equation}
    \label{tt8}
    V_{cav}(r;\lambda) = A \ e^{-r/B+\lambda} \ . 
\end{equation}
Note that, by appropriate tunings of the  $\lambda$ parameter,  one can move from a practically non interactive solute, for $\lambda=\lambda_1$ sufficiently large and negative, up to the limit in which the solute expels solvent atoms up to distances similar to the ones observed with the actual $V_{\rm He-\alpha}$ interaction, for more positive values of $\lambda=\lambda_2$. 
Of course, the actual choice of the values of these two parameters is strongly correlated with the value of $A$ .

\begin{figure}[h!]
        \centering
        \includegraphics[width=1\textwidth]{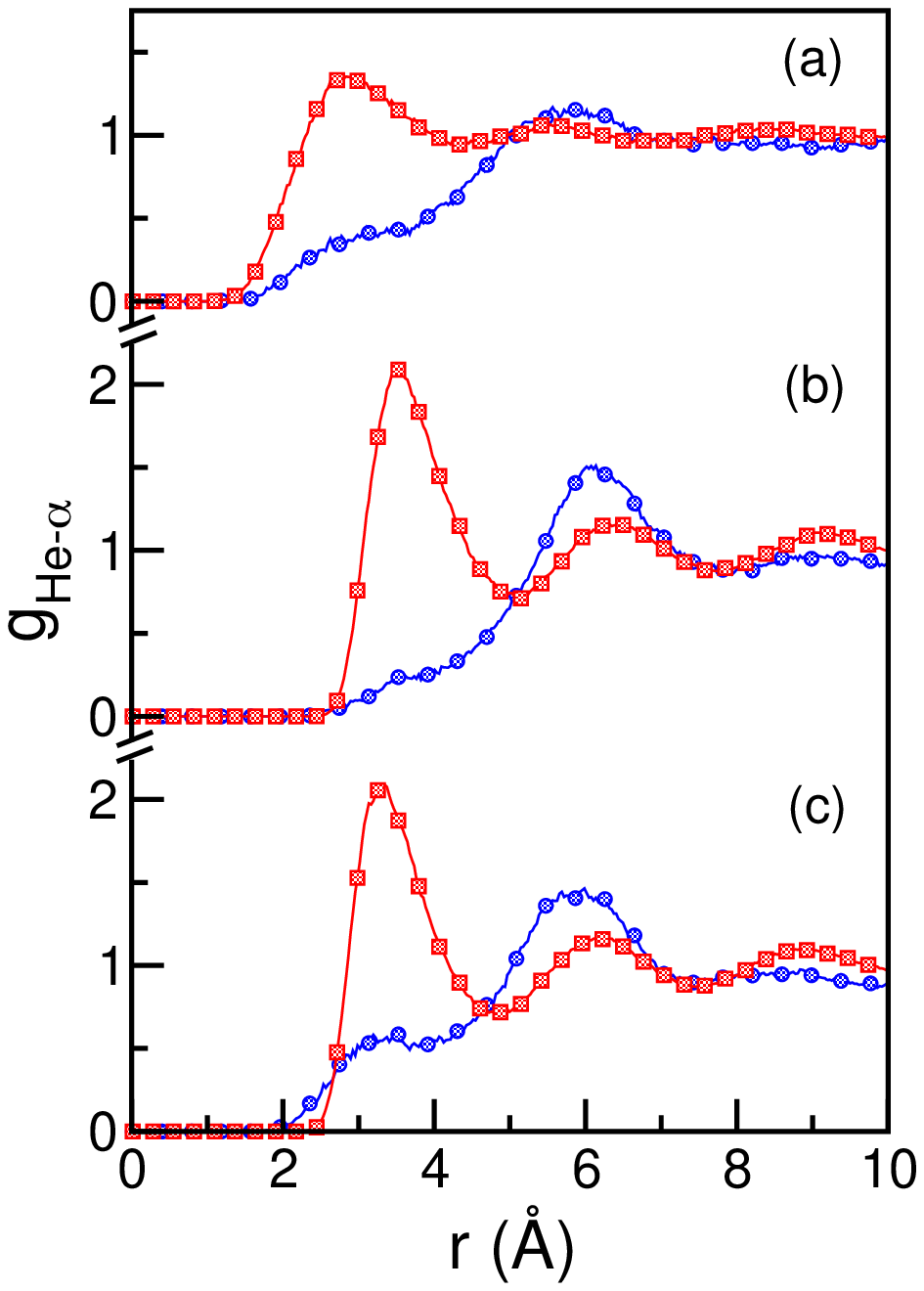}
        \caption{He-solvent pair correlation functions for the final states of the different stages of the solute incorporation within the eutectic Li-Pb mixture. Li: blue circles; Pb: red squares. (a): growth; (b):  insertion; (c): shrinkage (see text).}
        \label{fig:gdr_HeS}
\end{figure}

Along the second stage, hereafter referred to as the {\it insertion} stage, changes in the chemical potential are computed by gradually adding an extra linear term of the form $\lambda V_{\rm He-X}$,
so that the resulting solute-solvent coupling is now of the form:
\begin{equation}
V_{ins}(r;\lambda) = V_{cav}(r;\lambda_2)+\lambda \ V_{\rm He-X}(r)
\end{equation}
with $\lambda$ varying between 0 and 1. 
Along the third and final {\it shrinkage} stage,
the actual solute-solvent interaction is recovered by 
switching off the cavity term. To do so one has to  
reversibly reduce the value of $\lambda_2$ to its original $\lambda_1$ value. 

Collecting all that, the excess chemical potential can be computed as:
\begin{eqnarray}
\mu_{ex}&=&\mu_{gro}+\mu_{ins}+\mu_{shr} \nonumber \\
&=&\int_{\lambda_1}^{\lambda_2}{\rm d}\lambda \
\langle V_{cav}(\lambda)\rangle_{V_{cav}(\lambda)}
+
\int_0^1
{\rm d}\lambda \
\langle V_{\rm He-\alpha}\rangle_{V
_{cav}(\lambda_2)+\lambda  V_{\rm He-X}
}
+\nonumber \\
&& \hspace{4.2cm}+
\int_{\lambda_2}^{\lambda_1}
{\rm d}\lambda \
\langle V_{cav}(\lambda)\rangle_{
V_{cav}(\lambda)+V_{{\rm He-\alpha}}
} 
 \ \ .
 \label{eq9}
\end{eqnarray}
In all cases,  we have found reasonable results by setting $A=12.48$ eV and $B=1$~\AA$^{-1}$ in Eq.~(\ref{tt8}), and $\lambda_1=-8$ and $\lambda_2=-2$ in Eq.~(\ref{eq9}).

The three average values that appear in the integrands in Eq.~(\ref{eq9}) can be cast in terms of the solute-solvent pair correlation functions, i.e.:

\begin{equation} \label{avg_V}
    \langle{V}\rangle_{\lambda, \xi} = 
    4\pi\sum_{\alpha} \rho_\alpha \int_0^\infty {\rm d}r \ r^2 \ g_{\rm He-\alpha}^{\lambda, \xi}(r) V_{\rm He-\alpha}(r) \ \ ,
\end{equation}
where

\begin{equation}\label{gdr}
    g^{\lambda, \xi}_{\rm He-\alpha}(r) = \frac{1}{4\pi\rho_\alpha}\sum_{i=1}^{N_\alpha}
    \langle \delta(|{\bf r}_{\rm He}-{\bf r}_i^\alpha|-r)\rangle_{\lambda, \xi}. \ \ 
\end{equation}
In Eqs.~(\ref{avg_V}) and (\ref{gdr}) $\lambda, \xi$ denote the stage at which the pair correlations functions are calculated, with corresponding coupled Hamiltonian ${\cal H} = {\cal H}_0 + V(\lambda, \xi) = {\cal H}_0 + V_{cav}(\lambda) + \xi V_{\rm He-\alpha}$. 

\clearpage
\section{Results and discussion}
\label{results}
\subsection{Eutectic mixture}
We start discussing the simulation predictions for the solubility of He in the eutectic mixture at 508~K. 
We find that at 508 K the average global density at 1~bar is $\langle \rho_S\rangle = 0.0338$~\AA$^{-3}$, a value that lies within the uncertainty reported in the literature for the experimental results, i.e. $\rho_{exp} = 0.0334$ \AA$^{-3}$\cite{khairulin2017volumetric,saar1987calculation} and $0.0341$ \AA$^{-3}$.\cite{de2008lead} 
Regarding the excess chemical potential, we find $\mu_{ex}=16.2 \  k_BT$ which,  translated into solubility values, yields 
$(5.0 \pm 1.1)\times10^{-16}$~Pa.  In Table~\ref{tab:tableI} we show the three-stage contributions to the excess chemical
potential yielding $\beta \mu_{ex} = 16.2$.  As expected, the leading result is dominated by the initial {\it growth} procedure, along 
which the major rearrangement in the local structure of the solvent around the solute takes place.  The subsequent {\it insertion} stage adds an additional positive contribution to the chemical potential as a result of the increment in the repulsive interactions provided by the gradual turning on the actual solute-solvent potential. Finally, the reverse trend is verified along the third stage, where the reduction in the chemical potential goes hand in hand with the gradual cancelation of the initial repulsive contribution provided by $V_{cav}$. 

Analysis of the different solute-solvent pair correlation functions, shown in Fig.~\ref{fig:gdr_HeS}, provides microscopic descriptions of the solvent structures at the end points of each stage.  The solvation structure after the first stage is characterized by a somewhat wide first solvation shell located between 2~\AA \ and 4~\AA \ away from the solute. The second stage produces a $\sim 1$~\AA \ narrowing of the closest shell, which shows a more ordered structure.  Finally, the clearest modification observed after the {\it shrinkage} process involves a mild change in the Li profile. 

We remark that, in all cases, the differences between the magnitudes of the main peaks of the pair correlation functions reveal clear preferential solvation of Pb (with respect to the uncorrelated, ideal behavior), in detriment of Li.  Interestingly, such trend reverses in the second solvation shell located at $\sim 6$~\AA. As such, our results indicate that the solvation environment for the noble gas can be cast in terms of a cavity surrounded exclusively by Pb atoms, which contrasts with an excess population of Li atoms in the second solvation shell. From energy grounds, this observation is corroborated by decomposing the excess chemical potential into contributions from the two solvents.  The results are listed in the second and third rows of Table~\ref{tab:tableI}, and reveal a clear preponderance of the contributions of Pb along the three stages that involve the accommodation of the solute in the eutectic solution. 

\begin{table}[]
            \centering
            \begin{tabular}{ccccc}
            \hline
                 ~     & $\beta\mu_{gro}$ & $\beta\mu_{ins}$ & $\beta\mu_{shr}$ & $\beta\mu_{ex}$ \\ \hline
                 Total & 27.7 & 6.9 & -18.4 & 16.2\\ 
                 Pb    & 24.4 & 6.6 & -16.8 & 14.2\\
                 Li    & 3.3  & 0.3 & -1.6  & 2.0 \\
                 \hline                 
            \end{tabular}
            \caption{Excess chemical potentials of Li in the eutectic Li-Pb solution after the 3 stages. Also shown are results for the individual solvent contributions. }
            \label{tab:tableI}
        \end{table}

\subsection{Temperature and composition dependence of $K_H$}

    \begin{figure}
        \centering
        \includegraphics[width=1\textwidth]{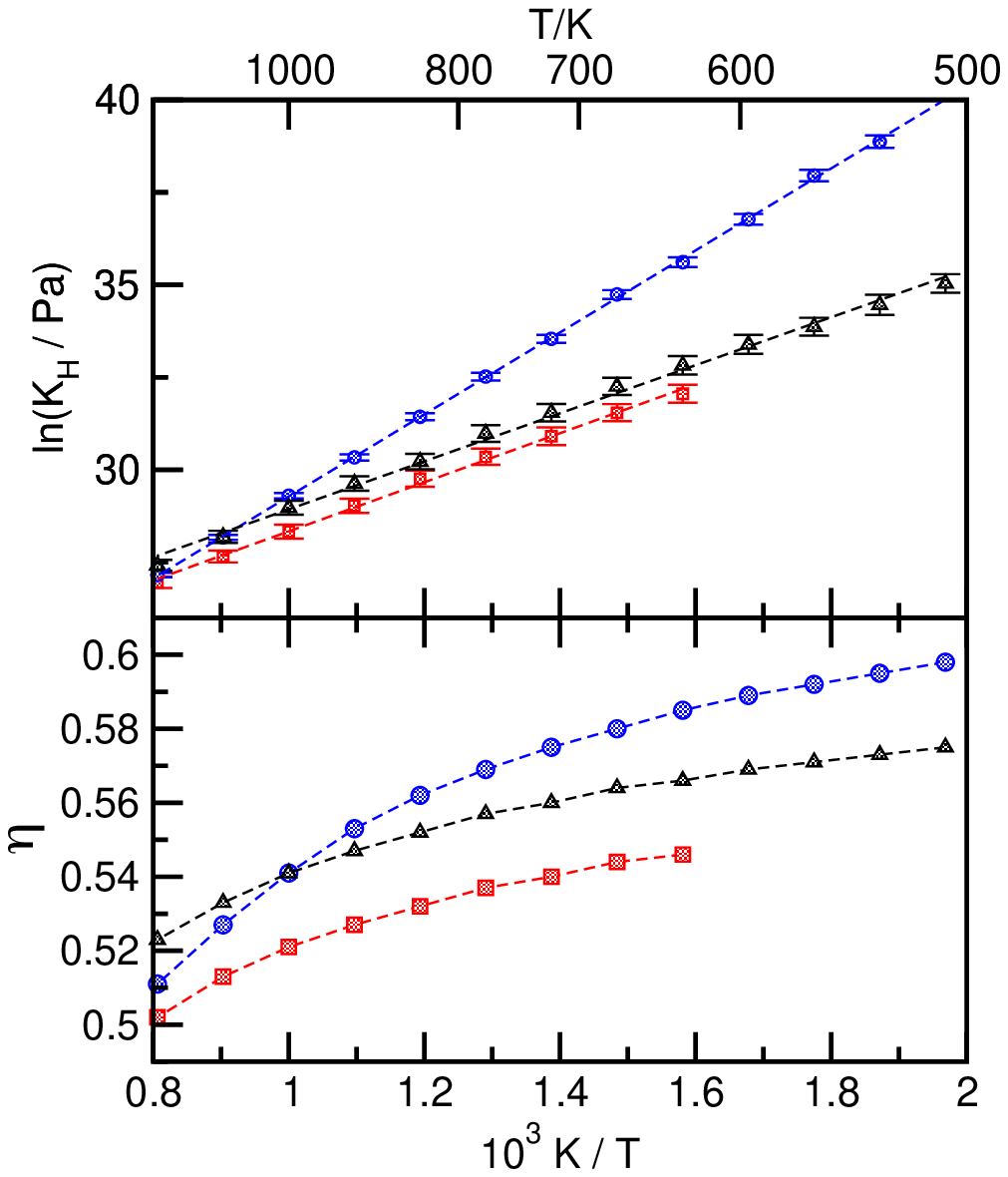}
        \caption{ Top panel:
        Temperature dependence of the Henry's constant for liquid mixtures at the eutectic concentration  (black triangles). Also shown are results for the pure Li (blue circles) and pure Pb (red squares). The lines represent linear fits of each set of data. Bottom panel: packing factors for the mixtures shown in the top panel (see text).
        } 
        \label{kh_vs_t}
    \end{figure}
  
The next aspect studied concerns the modifications operated in the He solubility in the eutetic mixture as the temperature is raised.
Results for $K_H(T)$ are depicted in the top panel of Fig.~\ref{kh_vs_t} (black triangles).  Note that, by doubling the temperature, the solubility increases two orders of magnitude. The comparison with similar temperature modifications for the case of the pure solvent is instructive. The results also appear in the same figure with red squares (Pb) and blue circles (Li).  As expected, the results for the eutectic mixture closely follow those of pure Pb, for temperatures above the melting point of this metal. 

The contribution of Pb takes over the one of Li because of two main reasons: (i) there is much more Pb than Li for eutectic mixtures; (ii) the first shell of atoms around He is mainly composed of Pb, while Li atoms are repelled to outer shells, as discussed above.
One may realize that He is more soluble in Pb that in Li at any temperature, given the smaller values of Henry's constants. %Esto quizá mover a discussion / concluding remarks

In contrast, the modifications on the solubility of He in pure Li resulting from the temperature changes are much more dramatic since, in the latter case,  Henry's constant when going from $T\sim 500$~K up to $T\sim 1000$~K is two orders of magnitude larger.  However, the differences in Henry's constant between different solvents in the high-temperature regime appear much more modest, an observation that agrees with previous results shown in Fig.~{B.8} of Ref.~\cite{alvarez2024henry} for different melts at similar temperatures. 

Next, we analyze the dependence of Henry's constant on the relative concentration of each component in the mixture, along the $T \sim 1000$~K isotherm, as it is very close to the lowest temperature at which all Pb-Li mixtures along the complete phase diagram present liquid-like characteristics.\cite{okamoto1993li}

\begin{figure}
            \centering
            \includegraphics[width=1\textwidth]{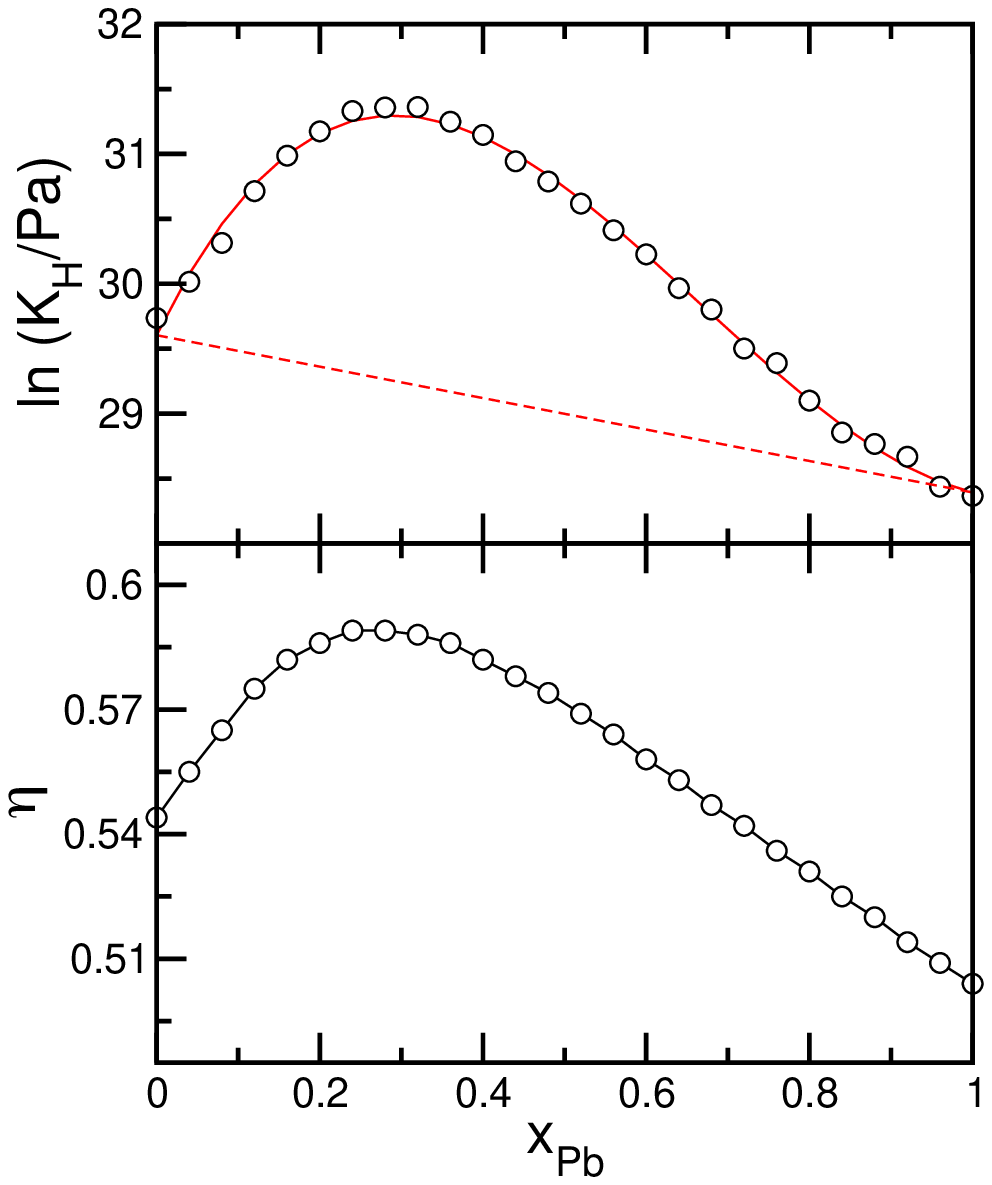}
            \caption{
            Top panel: Values of $K_H$ for He in different Li-Pb mixtures along the $T=1000$~K isotherm. Bottom panel: packing factor of the Li-Pb mixtures along the $T=1000$~K isotherm. The lines are an aid to the eye. The dashed line corresponds to the ideal mixing rule (see text).        
            } 
            \label{kh_eta_x}
        \end{figure}

Results for the Henry's constant as a function of the Pb molar fraction are displayed in the top panel of Fig.~\ref{kh_eta_x}. 
Clearly, the plot reveals a non-monotonic behavior of $K_H$, with a minimum of the solubility for $x_{\rm Pb}\sim 0.3$. 
It is worth mentioning that results for Henry's constants deviate -- especially in that low Li-concentration regime -- considerably from those defined by the following ideal mixing rule (depicted in the top panel of Fig.~\ref{kh_eta_x} with dashed lines):
\begin{equation} \label{KH_mix}
    \ln{K_{\rm H}(x)} = (1-x) \ln{K_{\rm H}(0)} + x \ln{K_{\rm H}(1)}  \; ,
\end{equation}
where $x=N_{\rm Pb} / (N_{\rm Li}+N_{\rm Pb})$ is the atomic fraction of Pb, while $K_{\rm H}(0)$ and $K_{\rm H}(1)$ correspond to the Henry's constants simulated for pure Li and Pb, respectively.

\begin{figure}
    \centering
    \includegraphics[width=1\textwidth]{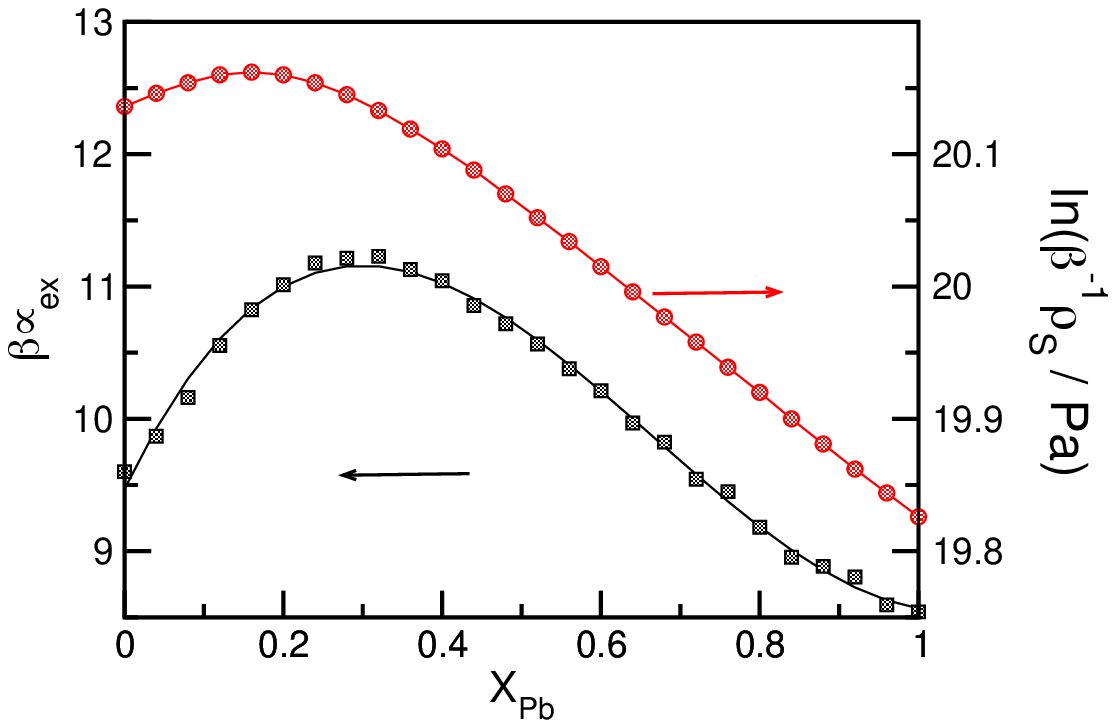}
    \caption{Excess chemical potential (left $y$-axis, black squares) and solvent global density (right $y$-axis, red circles) as a function of the atomic fraction of Pb along the $T=1000$ K isotherm.
    } 
    \label{mu_rho_x}
\end{figure}

\begin{figure}

        \centering
        \includegraphics[width=1\textwidth]{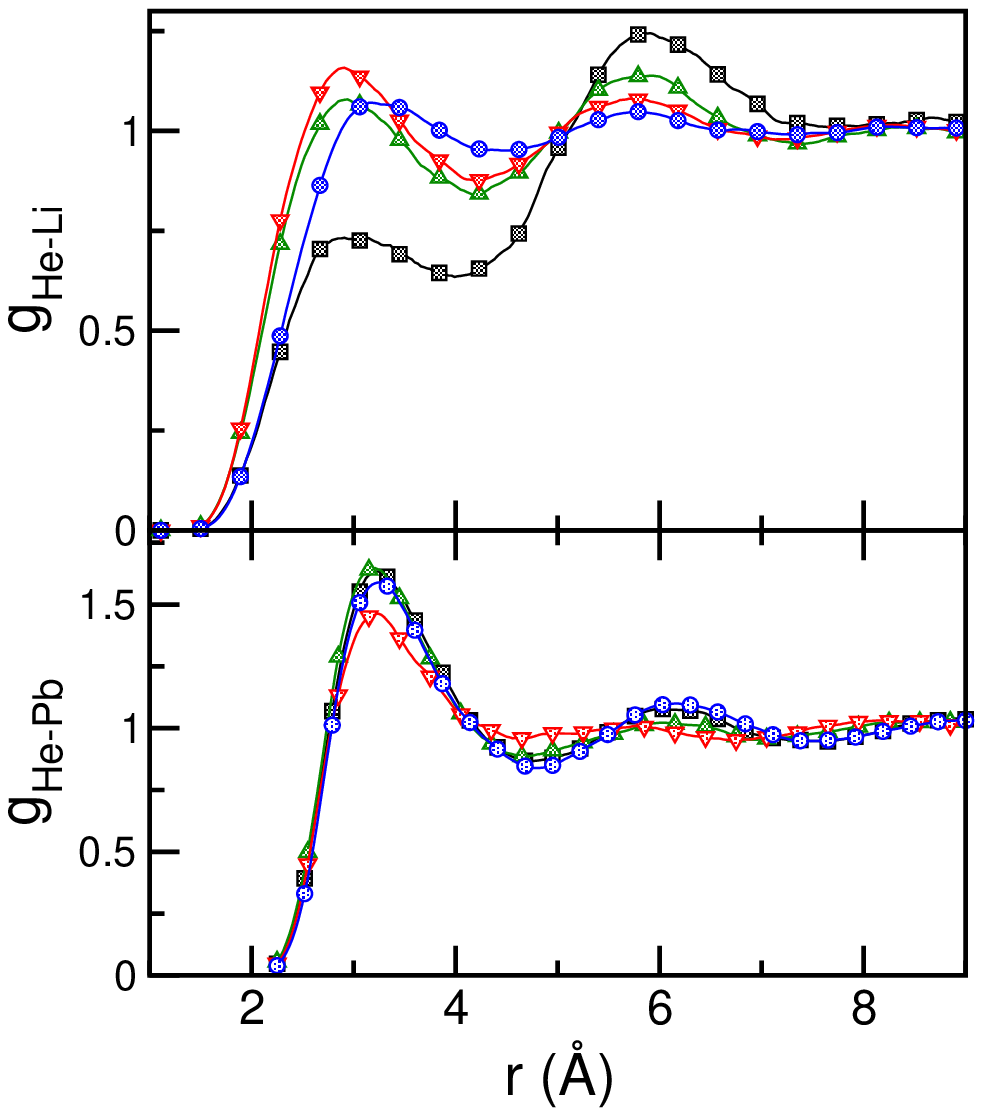}
        \caption{He-solvent pair correlation functions for different compositions of Li-Pb mixtures at $T=1000$ K. 
        $x_{\rm Pb}=0$ (blue circles, upper plot); $x_{\rm Pb}=0.28$ (red down-triangles); $x_{\rm Pb}=0.5$ (green up-triangles); 
        $x_{\rm Pb}=0.84$ (black squares); $x_{\rm Pb}=1$ (blue circles, lower plot).
        } 
        \label{gr_vs_x_1000K}
\end{figure}

Additionally, Fig. \ref{mu_rho_x}, shows that the two terms contributing to $\ln(K_H)$ in Eq.~(\ref{kh}) present a similar non-monotonicity, being the chemical potential the dominant one. 
Interestingly, direct inspection of the plots of the solute-solvent pair correlation functions at different compositions shown in Fig. \ref{gr_vs_x_1000K} reveals that the minimum in solubility correlates with a solvation structure in which the excess population of Li atoms in the first solvation shell -- i.e. with respect to the trivial, $g(r)=1$,  ideal result -- attains a maximum.  

Looking for additional clues to understand this behavior, given the essentially repulsive character of the solute-solvent interactions described in Sec.~\ref{model}, 
we focus on the distribution of voids that would serve to accommodate the solute previous to its insertion. 
A crude characterization of the such voids is given by the packing factors of the mixtures, given by:
\begin{equation}
    \eta = \frac{\pi}{6}\sum_\alpha \rho_\alpha \ \sigma_\alpha^3
    \ \ \ ,
\end{equation}
where the values of $\sigma_\alpha$ stand for the distances at which the corresponding pure solvent pair correlation functions present their first and main maximum, i.e., peak. 
In a crude hard-sphere model, $\eta$ represents the fraction of the total non-available volume for insertion of the guest solute. Results for the dependence of $\eta$ on the Pb concentration appear in the lower panel of Fig.~\ref{kh_eta_x}. 
One can readily observe a clear correlation with the minimum in solubility and the maximum value of the packing factor, a result that would corroborate that the values of $K_H$ for He in the liquid metal mixtures can be interpreted invoking the repulsive, hard-sphere-like characteristics of the solute-solvent interactions. 

\newpage
\section{Concluding remarks}
\label{conclusions}

In the present work we provide data from molecular dynamics simulations reporting the solubility of infinite diluted helium
in the lead-lithium eutectic mixtures, as relevant data for the future design of nuclear fusion reactors.  We have employed
a  classical Kirkwood perturbative approach in order to compute Henry's constants of helium in lithium-lead mixtures in
a wide range of temperatures and concentrations.  In all cases, we found that Henry's constants decrease (solubility increases) 
with increasing temperature, i.e. all values of the linear regression of $K_{\rm H}$ grow exponentially with ($1/T$). 

In systems with harshly repulsive interactions, the solubility behavior resembles that of a hard-sphere system, where the solubility is mainly influenced by the space occupied by solvent atoms. These atoms rearrange themselves to maximize the solubility of guest atoms.  All solvents exhibit an Arrhenius-like relationship for the temperature dependence of Henry's constants. This relationship indicates that while the density-dependent term behaves essentially as a constant, the chemical potential term shows more 
significant variations.

In contrast to predictions by Sedano et al.~\cite{sedano2022solubility}, who used a semi-empirical approach suggesting that eutectic mixtures would have smaller solubility values than pure solvents, our results align with hard-sphere models, which suggest that solubility is determined by the packing density of solvent atoms.  Specifically,  Pb-rich solvents show greater solubility for He than Li-rich solvents because Pb atoms are less densely packed than Li atoms due to their larger size.  Additionally, while ideal mixing rules do not apply to the Henry's constant of Pb-Li mixtures, the eutectic mixture’s solubility falls between those of the pure solvents. Our MD simulations also show that He atoms preferentially interact with Pb over Li, indicating density fluctuations around He differ between the two metals.

The maximum values for both Henry's constant and packing factor are consistent with experimental observations, such as the
divergence of heat capacities and the thermal expansion coefficient at around 1000 K and 22\% Pb  concentration,\cite{khairulin2017volumetric,saar1987calculation} which corresponds to a critical point in the phase diagram.\cite{okamoto1993li} From a reactor design perspective, incorporating Pb into breeding blankets not only helps to reduce reactivity but also significantly enhances the solubility of produced He, potentially by two orders of magnitude near the eutectic point.

\section*{Acknowledgments}
J.M. and L.B. acknowledge financial support from the EUROfusion project (HORIZON-101052200-EUROfusion).  J.M.  and E.A.  thank financial support of project PID2021-124297NB-C32 funded by MCIN/AEI/10.13039/5011000-11033 and ERDF “A way of making Europe” by the “European Union NextGenerationEU/PRTR”.  F.M. and J.M.  gratefully acknowledge financial support from the Generalitat de Catalunya (Grant 2021 SGR 01411).
F.M. acknowledges support under project PID2023-147469NB-C21 by the Ministerio de Ciencia e Innovaci\'on MCIN/AEI/10.13039/501100011033 (Spain).
E.A.  has been awarded a fellowship at the Polytechnic University of Catalonia by the Spanish "Consejo de Seguridad Nuclear" through the {\it ARGOS} Chair of Nuclear Safety and Radiation Protection. 

\bibliographystyle{elsarticle-num} 
\bibliography{references}

\color{black}
\clearpage
\appendix
\section{New parametrization of the EAM of Li}

\begin{figure}[t]
    \centering
    \includegraphics[width=0.9\linewidth]{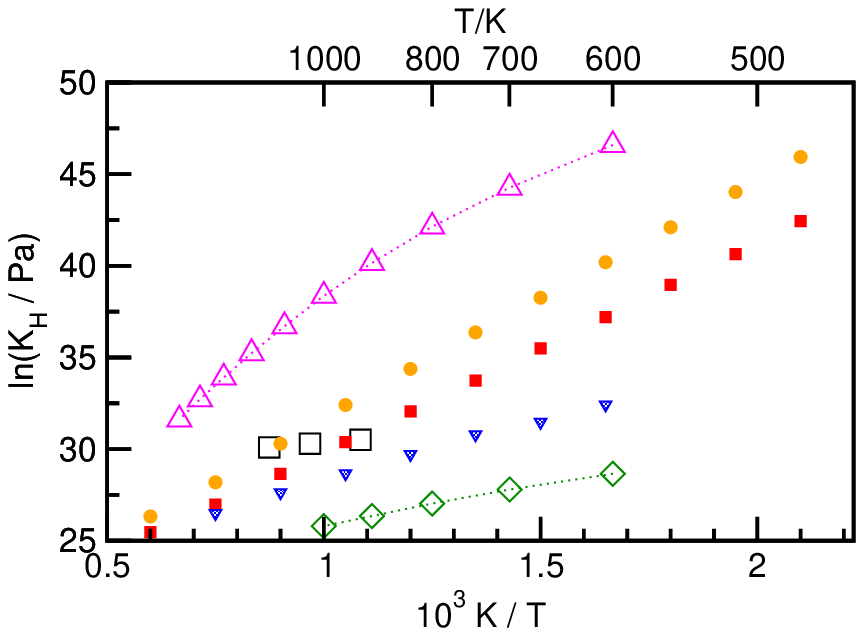}
    \caption{Temperature dependence of the logarithm of Henry's constants for simulations using different pure solvent models: Al-Awad's Li\cite{al2023parametrization} (orange circles), Belashchenko's Li\cite{belashchenko2012embedded} (red squares) and Belaschenko's Pb\cite{belashchenko2012computer} (blue triangles). Khairulin's et al. model predicts the results depicted with pink triangles and green diamonds, for Li and Pb (respectively). Experimental values for Li are represented with black squares.}
    \label{fig:KH_vs_T_pure}
\end{figure}

In contrast to our previous work for alkali metals,\cite{alvarez2024henry} in this work we use a new parametrization of the Li potential performed by Al-Awad et al,\cite{al2023parametrization} which has been shown to perform better estimations Pb-Li crossed RDFs rather than the original Belashchenko's EAM for Li\cite{belashchenko2012embedded} when mixed with Belashchenko's EAM for Pb \cite{belashchenko2012computer} as used in Ref.~\cite{belashchenko2019inclusion}.

In Fig.~\ref{fig:KH_vs_T_pure} we compare the logarithms of Henry's constants of both Al-Awad's and Belashchenko's EAM. Experimental data for Li\cite{slotnick1965solubility} (squares) lie in between the two predictions. So, in terms of solubility, both models seem to agree in a way similar to experiments.

The solubility of He in Pb has not been studied experimentally to date,
to the best of our knowledge. It was studied by Shpil'rain et al.\cite{shpil2007solubility} using a ``hole" model.
However, the latter underestimates by several orders of magnitude the solubility of He in pure Li compared to both experimental data and our calculations. 
In the case of pure Pb, where no experimental data have been found, it seems that the solubility of He in pure Pb is, in the opposite way, overestimated.
The model captures an increasing trend of the solubility with temperature in 
both cases and also that the solubility of He in Li is smaller than in Pb.

\section{Size independence of the results}

\begin{figure}[t]
    \centering
    \includegraphics[width=0.9\linewidth]{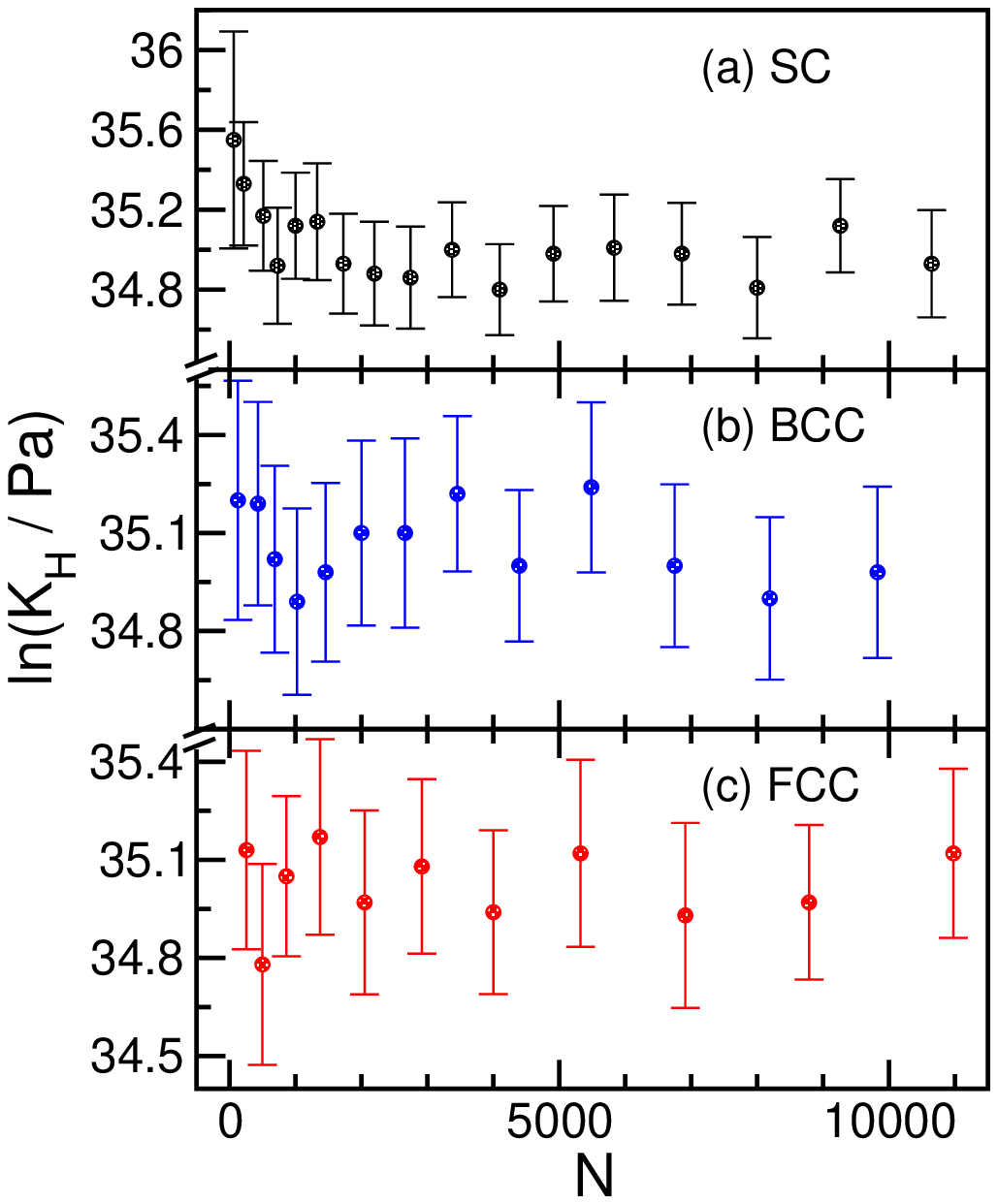}
    \caption{Logarithm of Henry's constants at the eutectic point (508~K and 16\%Li-84\%Pb) for reference LM systems that have been melted from initial lattices of type: (a) SC-lattice (black circles); (b) BCC-lattice (blue circles); (c) FCC-lattice (red circles).}
    \label{fig:convergence}
\end{figure}

There are two main reasons for the choice of $N=1024$:
\begin{enumerate}
    \item It allows us to initialize the system by melting a body-centered-cubic (BCC) lattice occupying the whole box.
    \item We have verified that larger simulation boxes do not improve the precision of the calculations -- since boxes have to be larger than the cutoff radius of the He-LM interactions, i.e., large enough to avoid finite-size effects -- but much larger boxes increase the computational cost. 
\end{enumerate}
In order to illustrate the second reason, we have run extra simulations departing from simple-cubic (SC), BCC and face-centered-cubic (FCC) lattices with total amounts of $n^3$, $2 n^3$ and $4 n^3$ LM atoms, respectively, where $n$ is a positive integer number.
In Fig.~\ref{fig:convergence} we see that calculations departing from any of the considered lattices, for $N \gtrsim 1000$, yield similar results, within the computed uncertainties.

\end{document}